\documentclass[]{wechat}

\usepackage{amsmath,amsfonts,bm}

\def\eqref#1{equation~\ref{#1}}

\def\1{\bm{1}}

\DeclareMathAlphabet{\mathsfit}{\encodingdefault}{\sfdefault}{m}{sl}
\SetMathAlphabet{\mathsfit}{bold}{\encodingdefault}{\sfdefault}{bx}{n}

\usepackage{url}
\usepackage{threeparttable}
\usepackage[table]{xcolor} 
\usepackage[bottom]{footmisc} 

\usepackage{CJKutf8}

\usepackage{amsmath,amssymb}
\usepackage{booktabs}
\usepackage{tabularx}
\usepackage{array}
\usepackage{graphicx}
\usepackage{xcolor}
\usepackage{bbm}
\usepackage{algpseudocode}
\usepackage{algorithm}
\usepackage{graphicx}

\graphicspath{{figure/figs/}}

\newcolumntype{Y}{>{\centering\arraybackslash}X}

\title{WeSearch: A Unified Model-Centric Retrieval Framework for WeChat Search} 
\author{%
\fontsize{9.2pt}{11.2pt}\selectfont
   Kepu Zhang$^{*}$, Yuzhe Cai$^{*}$, Yue Shu$^{*\dagger}$, Dawei Chen, Jinhai Zhu, Yuan Xu, Zehua Dai, Tongke Ni, Ziyao Chen, \\Chen Fu, Jie Shao, Chao Wang, Xiangmin Wang, Ruifeng Zhao, Rongchun Yao, Jie Yin, Weixin Qiu, \\Nan Song, Junyang Li, Fan Li, Quanchao Hui, Yulong Ou, Qiang Chen, Hao Cao, Sen Yang \\
\vspace{6pt}
WeChat Search Group, Tencent Inc.\\
\vspace{4pt}
\vspace{-24pt}
}

\vspace{-15pt}
\abstract{
We present \textbf{WeSearch}, a unified, model-centric retrieval framework for large-scale, continuously evolving content search. Prevailing methods assign documents fixed hierarchical Semantic IDs, implicitly assuming each
document follows a single coherent semantic path---an assumption that often breaks down for multi-topic documents. WeSearch instead retrieves query-relevant clusters and then ranks atomic documents within the resulting
query-specific candidate space, decoupling candidate pruning from fine-grained relevance modeling. Combined with relevance-aware training, compact document representations, and incremental index updates, this design
enables weekly---or daily when needed---model refreshes. Deployed as the primary semantic retrieval path in WeChat Search alongside a lexical syntax tree, WeSearch shows consistent one-week A/B gains on corpora from hundreds
of millions to billions of documents, and a one-month full-traffic A/B test confirms stable improvements of $+0.68\%$ CTR and $+2.27\%$ per-user article clicks, together with a $41\%$ reduction in retrieval resource cost and a $19$\,ms reduction in graphic-text retrieval latency. WeSearch is now fully deployed in WeChat Search, serving $100\%$ of production traffic.
\vspace{-3.5pt}
}
\date{\today}

\begin{document}
\maketitle

\begingroup
\renewcommand{\thefootnote}{}
\footnotetext{\scriptsize
$^{*}$Equal contribution (co-first authors).
\quad
$^{\dagger}$Project leader and corresponding author: \url{vincentshu@tencent.com}.
}
\endgroup

\vspace{-22pt}
\begin{figure}[!b]
\centering
\includegraphics[width=0.92\textwidth]{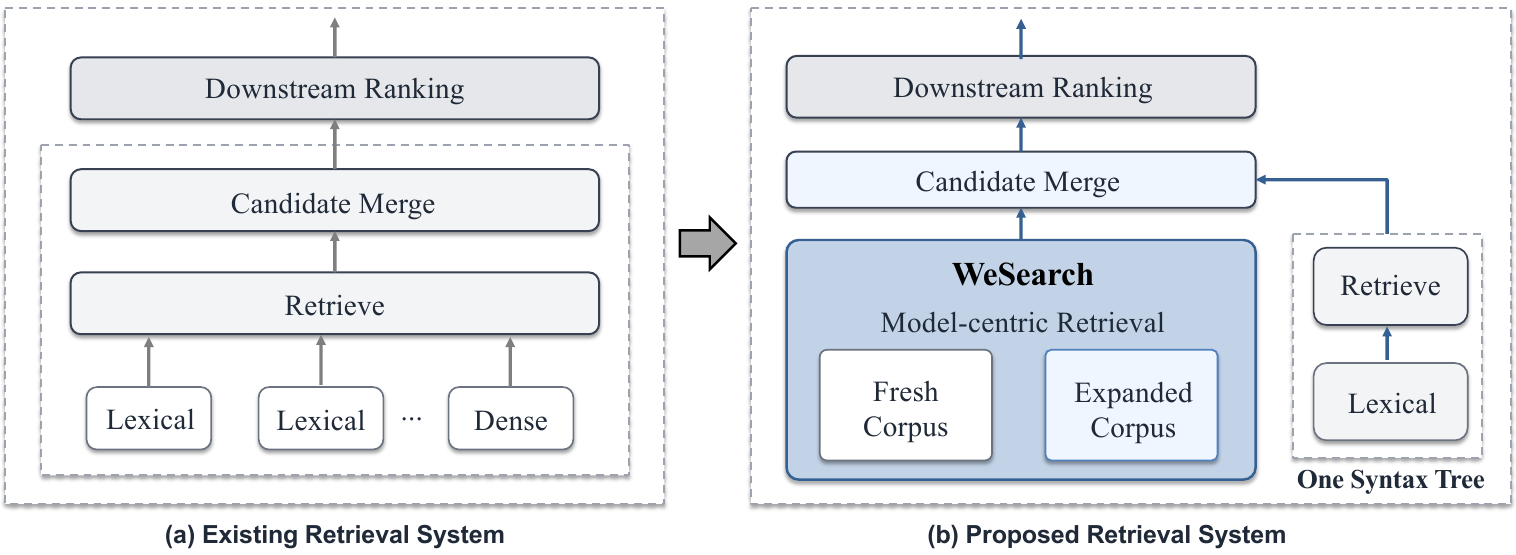}
\caption{\textbf{Comparison between the existing retrieval system and the proposed retrieval system.} (a) The existing system combines multiple lexical and dense retrieval sources. (b) The proposed system uses model-centric retrieval as the primary semantic retrieval path and retains one syntax tree as a complementary lexical retrieval source.}
\label{fig:wesearch_online_architecture}
\end{figure}

\section{Introduction}
\label{sec:introduction}

Large-scale content search must retrieve relevant documents from collections containing hundreds of millions or even billions of items, while meeting strict requirements on latency, resource consumption, and freshness. Modern industrial search systems, such as WeChat Search, typically combine inverted-index retrieval~\cite{zobel2006inverted} with dense vector retrieval~\cite{10.1145/276698.276876}. Inverted retrieval is efficient and interpretable, but its coverage is constrained by lexical matching, query rewriting, and intersection strategies. Dense retrieval provides stronger
semantic matching, yet its candidates may still be removed by downstream filtering, quota limits, or ranking stages. Maintaining multiple retrieval paths can improve coverage, but may also introduce duplicated computation, communication overhead, and fragmented optimization objectives.

Generative retrieval has emerged as an alternative paradigm for information retrieval, motivated in part by advances in large language models (LLMs)~\cite{li2025matching,zhu2025large}. Existing approaches generally either treat each document as an independent output symbol~\cite{NEURIPS2022_892840a6}, resulting in a large output space, or assign
documents fixed hierarchical Semantic IDs and decode them level by level~\cite{chen2025onesearch,deng2025onerec}. Although hierarchical Semantic IDs reduce the output space, they assume that each document can be represented by
a single coherent semantic path. This assumption can be problematic for content-oriented collections, where articles and graphic-text posts may span multiple entities, topics, and user intents. Moreover, an error at an early decoding level can exclude the target document from subsequent retrieval stages.

We present \textbf{WeSearch}, a model-centric, cluster-constrained retrieval framework for large-scale and continuously evolving content search. WeSearch uses clusters to reduce the query-specific candidate space, rather than as fixed hierarchical identifiers for documents. Given a query, it retrieves relevant clusters through vector similarity search, gathers their associated documents, and then performs fine-grained matching and ranking over the resulting atomic-document candidates. Unlike prior generative retrieval methods that bind each document to a single hierarchical semantic ID and decode it token by token—where an error at an early prefix permanently prunes all good documents under the correct branch—WeSearch retrieves a \emph{set} of relevant clusters and takes the union of their member documents as candidates, keeping the candidate pool as large and inclusive as possible. This design combines the efficiency of cluster-based candidate reduction with document-level relevance modeling, without requiring each document to follow a single fixed semantic path.

To support retrieval at industrial scale, WeSearch combines multi-source relevance supervision and hard-negative mining with OPQ-based document representations~\cite{ge2013optimized}, asymmetric distance computation, and distributed scoring optimizations. A rolling-mounting mechanism allows fresh documents to be added to the existing cluster space without rebuilding the entire index, while periodic rolling training helps the model adapt to evolving query--document relevance.
In the previous production pipeline, the retrieval embedding model was typically updated monthly, limiting how quickly the retrieval space could adapt to new content, emerging topics, and changes in user behavior. WeSearch supports coordinated model refreshes on a weekly cadence, with updates as frequent as daily when needed, enabling the system to incorporate model improvements and evolving relevance patterns more promptly.

We also redesign the online retrieval architecture around WeSearch. As shown in Figure~\ref{fig:wesearch_online_architecture}, WeSearch serves as the primary semantic retrieval path, complemented by a single syntax tree for lexical matching. Candidates from the two paths are merged and deduplicated before being passed to downstream ranking. This design consolidates semantic retrieval into a unified model-based path while retaining a lightweight lexical source for exact matching.

We evaluate WeSearch in the WeChat Search system across three corpus configurations: a \textit{Base} corpus containing hundreds of millions of documents, an \textit{Expanded} corpus containing billions of documents,
and the \textit{Expanded} corpus augmented with a \textit{Fresh} corpus of tens of millions of newly arriving documents. Controlled offline experiments on the Base corpus examine cluster granularity, hierarchical Semantic-ID-based
retrieval, and document quantization, and show that hierarchical Semantic-ID-based retrieval is less suitable for heterogeneous content documents, while OPQ-based representations consistently outperform compact low-dimensional dense representations. We further validate training-data mining, structured document features, and teacher-assessed result quality at large scale.

Online, one-week A/B tests across all three corpus configurations yield statistically significant improvements over the corresponding production baselines. On the Expanded corpus with Fresh corpus configuration, a subsequent one-month full-traffic A/B test shows stable gains of $+0.68\%$ in CTR and $+0.71\%$ in average dwell time on the search result page, together with a $+2.27\%$ increase in per-user article clicks. In the same deployment, WeSearch reduces overall retrieval resource cost by $41\%$ and graphic-text retrieval latency by $19$\,ms. Based on this combined evidence of quality improvement and efficiency gains, WeSearch has been fully deployed in WeChat Search and currently serves $100\%$ of production traffic.

Our main contributions are as follows:

\begin{itemize}
    \item We propose \textbf{WeSearch}, which uses clusters solely to reduce the query-specific candidate space and performs fine-grained relevance modeling directly over atomic documents, avoiding the single-path assumption imposed by fixed hierarchical Semantic IDs and thus better accommodating multi-topic documents.

    \item WeSearch demonstrates the practical deployment of a model-centric retrieval framework as the primary retrieval path for content-oriented search at billion-document scale. Its system design combines rolling training, OPQ-based document representations with ADC-accelerated scoring, and incremental document mounting to support regular model and index updates.

    \item We conduct controlled offline experiments, one-week online A/B tests across three corpus configurations ranging from hundreds of millions to billions of documents, a one-month full-traffic online A/B test, and pairwise human evaluation. The results consistently demonstrate retrieval-quality improvements, positive online gains, and reduced retrieval resource cost and latency, and WeSearch has been fully deployed in production WeChat Search.
\end{itemize}

\section{Related Work}

\subsection{Sparse and Dense Retrieval for Large-Scale Search}

Large-scale industrial search systems typically combine sparse retrieval based on inverted indexes with dense retrieval based on vector similarity. Sparse retrieval provides efficient lexical matching, whereas dense retrieval improves semantic coverage by representing queries and documents in a continuous embedding space.

Inverted-index-based sparse retrieval, exemplified by BM25~\cite{10.1561/1500000019, TREC3}, has long served as the backbone of text search, grounded in statistical frameworks such as inverse document frequency~\cite{106765.106782} and efficient term-to-document mappings~\cite{schutze2008introduction}. Modern extensions further improve term weighting and query expansion to mitigate vocabulary mismatch~\cite{inproceedings, formal2021splade}. Despite their efficiency, sparse methods struggle with semantic matching and cannot robustly handle paraphrased expressions or conceptual relevance when surface vocabulary differs.

Dense retrieval encodes queries and documents into continuous vectors and retrieves candidates by semantic similarity~\cite{karpukhin-etal-2020-dense, xiong2021approximate}. It is less sensitive to exact lexical overlap and better handles paraphrases and semantically related expressions, making it a core component in modern industrial search stacks. For billion-scale corpora under strict latency constraints, dense retrieval typically relies on approximate nearest neighbor (ANN) indexes~\cite{10.1145/276698.276876}. Graph-structured indexes such as NSW~\cite{MALKOV201461}, HNSW~\cite{10.1109/TPAMI.2018.2889473}, NSG~\cite{fu2025fastapproximatenearestneighbor}, and DiskANN~\cite{subramanya2019diskann} achieve high recall by greedily navigating proximity graphs. However, building and updating such indexes over dynamically evolving corpora is expensive, and query-time filtering or attribute constraints substantially degrade traversal efficiency~\cite{10.1145/3786635}.

At the representation level, a single embedding vector collapses heterogeneous document semantics into a fixed-dimensional space, which often fails to capture fine-grained query--document relations for multi-topic, multi-entity, or multimodal content~\cite{weller2026theoreticallimitationsembeddingbasedretrieval, 10.1145/3397271.3401075, 3737916.3741120}. A natural response is to model relevance at a finer granularity, but applying such models directly over a billion-scale corpus is computationally prohibitive. These limitations motivate two-stage designs that decouple candidate reduction from fine-grained relevance scoring, where clustering offers a natural way to perform lightweight, query-dependent pre-filtering before detailed scoring is applied over individual documents.

\subsection{Generative Retrieval}

Generative retrieval formulates information retrieval as a conditional generation problem: instead of retrieving documents via an explicit index, a generative model directly predicts document identifiers conditioned on the input query. This unifies query understanding and candidate generation within a single model and allows retrieval to benefit directly from advances in sequence modeling.

DSI~\cite{NEURIPS2022_892840a6} first proposes training transformer-based models to map queries to document identifiers, enabling retrieval through direct generation. Subsequent work has explored various aspects of this paradigm. TIGER~\cite{rajput2023recommender} uses residual quantized autoencoders to generate Semantic IDs and trains encoder-decoder models for sequential recommendation. DSI++~\cite{mehta2023dsi++} and IncDSI~\cite{kishore2023incdsi} address the challenge of adapting generative retrievers to dynamic corpora. More recently, industrial-scale systems including OneRec~\cite{deng2025onerec}, OneSearch~\cite{chen2025onesearch}, and MTGR~\cite{han2025mtgr} have demonstrated the practical viability of generative retrieval, optimizing its efficiency and scalability for real-world deployment.

Despite this progress, GR with hierarchical Semantic IDs rests on a strong structural assumption: documents sharing an early identifier prefix should form a coherent semantic group, and each document should be well represented by a single hierarchical path. This holds reasonably well for structured domains such as products, but is problematic for content-oriented documents~\cite{oussaidene2026pdmrpassagedrivenmultiiddocument, woldemariam2026exploringbottomupclusteringcreating}: articles and graphic-text posts frequently span multiple entities, topics, and user intents, so documents assigned to the same identifier node may have substantially different relevance to a given query.
A second, closely related issue is early-stage error propagation~\cite{lee-etal-2026-documents}: when identifiers are decoded step by step, an incorrect decision at an early step permanently prunes all good documents under the correct branch. The top-level decision must therefore simultaneously provide a compact semantic representation and preserve query-specific recall—two requirements that often conflict in heterogeneous content spaces.

WeSearch is motivated by these two limitations. Rather than committing each document to a single hierarchical path, WeSearch uses clusters only to reduce the query-specific candidate space, and performs fine-grained relevance scoring directly over atomic documents within the retrieved clusters. This removes the single-path assumption and avoids unrecoverable early-prefix errors, while retaining the efficiency benefit of cluster-level pruning.

\section{WeSearch: Two-Stage Model-Centric Retrieval}
\label{sec:wesearch}

\subsection{Overview}

\begin{figure}[t]
    \centering
    \includegraphics[width=\columnwidth]{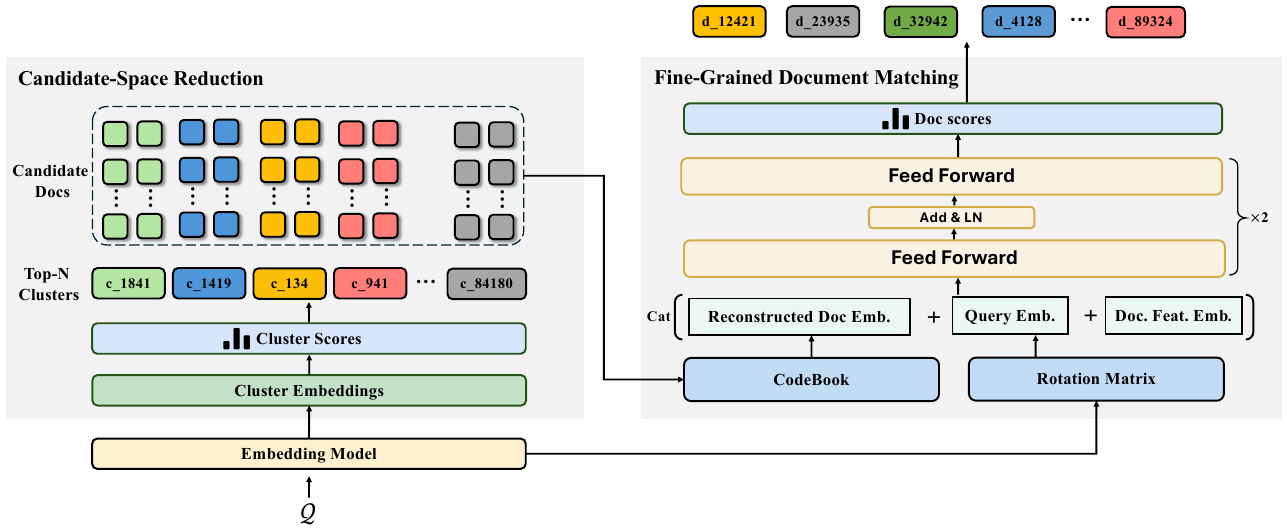}
    \caption{Overview of WeSearch. Clusters first define a query-specific candidate document space, over which fine-grained document-level ranking produces the final results.}
    \label{fig:wesearch_overview}
\end{figure}

WeSearch is designed for content-oriented search over a large and continuously evolving document collection $\mathcal{D}=\{d_1,\ldots,d_N\}$. Given a user query $q$, the system returns a ranked list of the top-$K$ documents:
\begin{equation}
    \mathcal{D}_{@K}^\prec = \mathcal{M}(q),
    \label{overall_equation}
\end{equation}
where $\mathcal{M}$ denotes the WeSearch model. As illustrated in Figure~\ref{fig:wesearch_overview}, $\mathcal{M}$ decomposes retrieval into two sequential stages: (i)~\textbf{cluster-level candidate-space reduction}, which uses clusters to narrow the full corpus down to a query-specific candidate set, and (ii)~\textbf{document-level ranking}, which performs fine-grained relevance scoring directly over atomic documents in that candidate set.

Unlike Semantic-ID-based retrieval~\cite{chen2025onesearch}, which assigns each document a complete hierarchical identifier and must decode it correctly end-to-end, WeSearch treats clusters purely as a candidate-space reducer: a cluster does not need to be a stable semantic label for all its member documents; it only needs to help form a query-specific candidate subspace with high recall. All fine-grained query--document discrimination is deferred to the document-level stage, where it operates directly on atomic documents.

Formally, let $\mathcal{C}=\{c_1,\ldots,c_M\}$ denote the cluster collection, with each cluster $c$ associated with a document subset $\mathcal{D}_c\subseteq\mathcal{D}$. In the \textbf{cluster stage}, WeSearch selects a query-dependent cluster set
\begin{equation}
\mathcal{C}_q
=
\operatorname{TopN}_{c\in\mathcal{C}}
s_c(q,c),
\label{eq:cluster_topn}
\end{equation}
where $s_c(q,c)$ is the cluster relevance score, and takes the union of their member documents as the candidate set:
\begin{equation}
\mathcal{D}_q
=
\bigcup_{c\in\mathcal{C}_q}\mathcal{D}_c.
\label{eq:candidate_documents}
\end{equation}

In the \textbf{document stage}, WeSearch ranks $\mathcal{D}_q$ to produce the final top-$K$ list. For efficiency, this stage is organized as a lightweight-to-heavy pipeline: a coarse scorer first shortlists $\mathcal{D}_q$ to the top-$P$ candidates $\mathcal{D}_q^P$ using compact semantic representations, and a fine-grained scorer then re-ranks $\mathcal{D}_q^P$ with richer document-side features $\mathbf{e}_f$:
\begin{equation}
\mathcal{D}_{@K}^\prec
=
\operatorname{TopK}_{d\in\mathcal{D}_q^P}
s_d(q, d, \mathbf{e}_f).
\label{eq:final_results}
\end{equation}

The cluster stage is recall-oriented, aiming to retain as many relevant documents as possible within $\mathcal{D}_q$, while the document stage is precision-oriented, performing fine-grained relevance discrimination over the retained candidates. The complete online inference procedure is summarized in Algorithm~\ref{alg:wesearch_inference}.

\begin{algorithm}
\caption{Online inference with WeSearch}
\label{alg:wesearch_inference}
\begin{algorithmic}[1]
    \Require Query $q$; cluster collection $\mathcal{C}$ with embeddings $\mathbf{E}_{\mathcal{C}}$; OPQ-coded document representations; document-side features
    \Ensure Top-$K$ retrieved documents $\mathcal{D}_{@K}^\prec$
    \State Encode query: $\mathbf{h}_q \gets f_{\theta}(q)$
    \Statex \textit{// Stage 1: Cluster-level candidate-space reduction}
    \State Retrieve relevant clusters: $\mathcal{C}_q \gets \operatorname{TopN}_{c\in\mathcal{C}} s_c(q,c)$
    \State Gather candidate documents: $\mathcal{D}_q \gets \bigcup_{c\in\mathcal{C}_q}\mathcal{D}_c$
    \Statex \textit{// Stage 2: Document-level ranking}
    \State Coarse shortlist by ADC-accelerated semantic similarity: $\mathcal{D}_q^P \gets \operatorname{TopP}_{d\in\mathcal{D}_q} s_{\text{coarse}}(q,d)$
    \State Reconstruct $\widehat{\mathbf{h}}_d$ from OPQ codes and gather $\mathbf{e}_f$ for all $d\in\mathcal{D}_q^P$
    \State Compute fine-grained scores $s_d(q,d,\mathbf{e}_f)$ for all $d\in\mathcal{D}_q^P$
    \State \Return $\mathcal{D}_{@K}^\prec \gets \operatorname{TopK}_{d\in\mathcal{D}_q^P} s_d(q,d,\mathbf{e}_f)$
\end{algorithmic}
\end{algorithm}

\subsection{Offline Index Construction}
\label{sec:wesearch_index}
\begin{figure}[t]
    \centering
    \includegraphics[width=0.9\columnwidth]{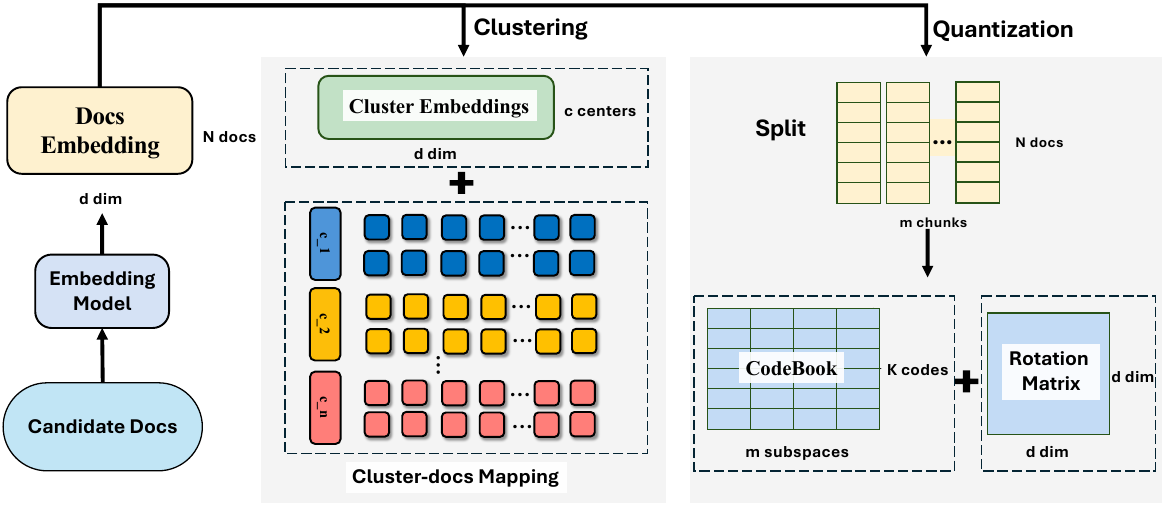}
    \caption{The offline index construction procedure.}
    \label{fig:offline}
\end{figure}

As shown in Figure~\ref{fig:offline}, offline index construction consists of two steps---\textbf{document clustering} and \textbf{document-representation compression}---and produces four artifacts that are directly loaded by the online retrieval service: (i)~a cluster embedding matrix $\mathbf{E}_{\mathcal{C}}$, (ii)~per-document cluster assignments $\{c_d\}$ defining the inverted mapping $\{\mathcal{D}_c\}$, (iii)~an OPQ rotation matrix $\mathbf{R}$ together with sub-space codebooks $\{\mathcal{B}_j\}_{j=1}^{m}$, and (iv)~per-document quantization codes $\{\mathbf{z}_d\}$. The first two support the cluster-level candidate-space reduction stage, while the latter two support the document-level ranking stage.

\subsubsection{Document Clustering}
\label{sec:wesearch_clustering}

Document clustering is the backbone of the cluster-level candidate-space reduction stage. Its purpose is not to recover a complete semantic taxonomy of the corpus, but to partition $\mathcal{D}$ into a tractable number
of candidate pools $\{\mathcal{D}_c\}$ that can be efficiently retrieved and unioned at query time. This objective substantially relaxes the demand for clustering precision: a cluster is useful as long as it concentrates
query-relevant documents, even if its member documents are not semantically homogeneous.

We adopt RQ-KMeans~\cite{luo2024qarmquantitativealignmentmultimodal}, a two-level residual-quantization-based clustering method, over all document embeddings. Compared with vanilla k-means, residual quantization recursively
clusters the residuals from previous levels, which yields a finer partition at comparable training cost---an important property for a billion-scale corpus.

Formally, given the dense representation $\mathbf{h}_d$ of each document $d \in \mathcal{D}$ and its two-level cluster assignment $c_d = (l_1,l_2)$, the document subset associated with cluster $c$ is
\begin{equation}
\mathcal{D}_c = \left\{ d \in \mathcal{D} \mid c_d = c \right\},
\label{eq:documents_in_cluster}
\end{equation}
and the embedding of cluster $c$ is the sum of its two-level codewords,
\begin{equation}
\mathbf{e}_c
= \mathbf{b}^{(1)}_{l_1} + \mathbf{b}^{(2)}_{l_2}.
\label{eq:cluster_embedding}
\end{equation}
All cluster embeddings are aggregated into the cluster matrix $\mathbf{E}_{\mathcal{C}} \in \mathbb{R}^{M \times D}$, which is loaded as a model parameter and used for cluster-level retrieval online.

\subsubsection{Document Representation Compression}
\label{sec:wesearch_compression}

Serving a billion-scale corpus requires that document representations fit entirely in memory while remaining accurate enough for fine-grained ranking. WeSearch adopts Optimized Product Quantization (OPQ)~\cite{ge2013optimized} to compress each document embedding into a compact code. Crucially, both the OPQ rotation matrix $\mathbf{R}$ and the sub-space codebooks $\{\mathcal{B}_j\}$ are instantiated as model
parameters, so that the quantization distortion is directly optimized against the downstream retrieval objective rather than being a post-hoc, retrieval-agnostic compression step.

Formally, we first apply the rotation
\begin{equation}
\widetilde{\mathbf{h}}_d = \mathbf{R}\,\mathbf{h}_d,
\label{eq:opq_rotation}
\end{equation}
and partition $\widetilde{\mathbf{h}}_d$ into $m$ sub-vectors
$\widetilde{\mathbf{h}}_d
= \big[\widetilde{\mathbf{h}}^{(1)}_d;\,\widetilde{\mathbf{h}}^{(2)}_d;\,\cdots;\,\widetilde{\mathbf{h}}^{(m)}_d\big]$.
For the $j$-th sub-space, a codebook
\begin{equation}
\mathcal{B}_j
=
\left\{\mathbf{b}_{j,1}, \mathbf{b}_{j,2}, \ldots, \mathbf{b}_{j,K}\right\}
\label{eq:codebook}
\end{equation}
is maintained, and the quantization code of $d$ in sub-space $j$ is obtained by selecting the nearest codebook vector,
\begin{equation}
z^{(j)}_d
= \arg\min_{k \in \{1,\ldots,K\}}
\big\|\widetilde{\mathbf{h}}^{(j)}_d - \mathbf{b}_{j,k}\big\|_2^2.
\label{eq:pq_code}
\end{equation}
The compressed representation of document $d$ is the concatenation of its sub-space codes,
\begin{equation}
\mathbf{z}_d
= \big[z^{(1)}_d,\, z^{(2)}_d,\, \ldots,\, z^{(m)}_d\big],
\label{eq:document_code}
\end{equation}
and the compressed representation reconstructed in the rotated space is
\begin{equation}
\widehat{\mathbf{h}}_d
= \big[\mathbf{b}_{1,z^{(1)}_d};\,\mathbf{b}_{2,z^{(2)}_d};\,\cdots;\,\mathbf{b}_{m,z^{(m)}_d}\big].
\label{eq:opq_reconstruct}
\end{equation}

Together, the clustering artifacts $\{\mathbf{E}_{\mathcal{C}}, \{c_d\}\}$ and the compression artifacts
$\{\mathbf{R}, \{\mathcal{B}_j\}, \{\mathbf{z}_d\}\}$ are packaged into the model checkpoint and shipped to the online service.

\subsection{Two-Stage Architecture}

The parameters of WeSearch consist of three parts:
(i)~a transformer-based query encoder $f_\theta$;
(ii)~offline-constructed index artifacts, including the cluster embedding matrix $\mathbf{E}_{\mathcal{C}}$, the OPQ rotation matrix $\mathbf{R}$, and the sub-space codebooks $\{\mathcal{B}_j\}$; and
(iii)~a lightweight feature-aware FFN ranker together with the embedding lookup tables for bucketed document-side features. The query encoder is initialized from a pretrained dual-tower retrieval model to provide a stable initial semantic geometry; the index artifacts are then constructed from the document representations produced by this initialized encoder (Sec.~\ref{sec:wesearch_index}) and stored as model parameters alongside the per-document OPQ codes $\{\mathbf{z}_d\}$; the ranker parameters are randomly initialized. 

\subsubsection{Stage I: Candidate-Space Reduction}

The first stage aims to rapidly filter the full corpus into a small, high-recall candidate subset at negligible online cost. Given a query $q$, the encoder produces a dense semantic embedding
\begin{equation}
    \mathbf{h}_q = f_{\theta}(q),
\end{equation}
which is scored against all cluster centroids using negative L2 distance:
\begin{equation}
    s_c(q, c) = -\left\| \mathbf{h}_q - \mathbf{e}_c \right\|_2.
\end{equation}
The top-$N$ clusters form the query-related cluster set $\mathcal{C}_q$, and their member documents are aggregated into the candidate document set $\mathcal{D}_q = \bigcup_{c \in \mathcal{C}_q} \mathcal{D}_c$. This stage tolerates coarse semantic grouping and prioritizes recall over fine-grained precision. Since it operates purely
on precomputed cluster centroids and requires no document-level computation, cluster selection compresses the search space from the billion-scale corpus to a tractable candidate pool.

\subsubsection{Stage II: Fine-Grained Document Matching}
\label{sec:wesearch_document_decoding}

The second stage performs fine-grained relevance discrimination over $\mathcal{D}_q$. Pure embedding distance is insufficient at this granularity: it does not reflect document quality, user behavioral signals, or temporal
relevance, all of which are critical for distinguishing semantically similar documents. Running a feature-enhanced nonlinear ranker over the entire $\mathcal{D}_q$, however, is too expensive to meet online latency
requirements. WeSearch therefore adopts a lightweight-to-heavy pipeline: (i)~an ADC-accelerated L2 coarse ranker shortlists $\mathcal{D}_q$ to a top-$P$ subset $\mathcal{D}_q^P$, and (ii)~a feature-aware FFN ranker produces
the final top-$K$ list over $\mathcal{D}_q^P$.

\paragraph{Coarse Ranking.}
To score every candidate in $\mathcal{D}_q$ efficiently, we project the query embedding into the OPQ-rotated space,
\begin{equation}
    \widetilde{\mathbf{h}}_q = \mathbf{R}\,\mathbf{h}_q,
\end{equation}
and compute negative L2 distance to each document using its OPQ codes $\mathbf{z}_d$ via asymmetric distance computation (ADC, Sec.~\ref{sec:wesearch_adc_system}):
\begin{equation}
    s_{\text{coarse}}(q, d)
    = -\left\| \widetilde{\mathbf{h}}_q
               - \widehat{\mathbf{h}}_d \right\|_2,
\end{equation}
where $\widehat{\mathbf{h}}_d$ is the rotated-space embedding reconstructed from $\mathbf{z}_d$ (Eq.~\ref{eq:opq_reconstruct}). The top-$P$ candidates are retained for fine-grained ranking:
\begin{equation}
    \mathcal{D}_q^P
    = \operatorname{TopP}_{d \in \mathcal{D}_q}\, s_{\text{coarse}}(q, d).
\end{equation}

\paragraph{Feature-Aware Fine-Grained Ranking.}
To complement the semantic signal, WeSearch incorporates multi-dimensional document-side features---including inherent document quality, user click and behavioral statistics, and temporal attributes---that capture practical
ranking clues beyond embedding distance. Each raw feature is bucketed into discrete levels and projected into a dense embedding through dedicated lookup tables that are stored as model parameters; their concatenation forms
the feature embedding $\mathbf{e}_f$ of a document.  Since only the OPQ codes $\{\mathbf{z}_d\}$ are stored in memory for the billion-scale corpus, the document side is represented by its rotated-space reconstruction
$\widehat{\mathbf{h}}_d$ (Eq.~\ref{eq:opq_reconstruct}) rather than the original embedding $\mathbf{h}_d$. A lightweight FFN then models nonlinear interactions among the query embedding, the reconstructed
document embedding, and the feature embedding:
\begin{equation}
    s_d(q, d, \mathbf{e}_f)
    = \mathrm{FFN}\!\left(
        \big[\, \widetilde{\mathbf{h}}_q;\;
                \widehat{\mathbf{h}}_d;\;
                \mathbf{e}_f \,\big]
      \right).
\end{equation}
The retained candidates are reordered by this score to produce the final retrieval result:
\begin{equation}
    \mathcal{D}_{@K}^\prec
    = \operatorname{TopK}_{d \in \mathcal{D}_q^P}\,
      s_d(q, d, \mathbf{e}_f).
    \label{eq:document_ranking}
\end{equation}
\section{Training WeSearch}
\label{sec:training}

WeSearch jointly learns two related but distinct retrieval capabilities. The first is \textbf{cluster coverage}: given a query, the model should select clusters that contain relevant documents. The second is \textbf{document discrimination}: given the documents collected from the selected clusters, the model should distinguish highly relevant documents from documents with lower relevance.

\subsection{Training Objective}
\label{sec:training_objective}

For a training query $q$, let $\mathcal{P}_q$ denote its set of positive documents and let $\mathcal{N}_q$ denote a set of negative documents. Each positive document $d^{+}\in\mathcal{P}_q$ is associated with one cluster that contains it. We denote one target cluster by $c^{*}$ and one target document by $d^{*}$.

\textbf{Cluster-level loss.}
The cluster-level objective encourages the query embedding to be geometrically close to the centroid of the cluster containing the relevant document: It is defined as a softmax cross-entropy loss over all cluster centroids:
\begin{equation}
\mathcal{L}_{\mathrm{cluster}}
=
-\log \frac{\exp\bigl(-\|\mathbf{h}_q - \mathbf{e}_{c^*}\|_2\bigr)}
{\sum_{c \in \mathcal{C}} \exp\bigl(-\|\mathbf{h}_q - \mathbf{e}_c\|_2\bigr)},
\label{eq:training_cluster_loss}
\end{equation}
where $\mathbf{e}_{c^*}$ is the centroid of the target cluster and the denominator sums over all clusters in $\mathcal{C}$.

\textbf{Document-level ranking loss.}
After constructing the query-specific candidate set $\mathcal{D}_q^P$ via ADC-accelerated coarse ranking, the document-level objective is a listwise cross-entropy loss that treats document ranking as a single-label classification problem over the top-$P$ candidates:
\begin{equation}
\mathcal{L}_{\mathrm{rank}}
=
-\log \frac{\exp\bigl(s_d(q, d^{*}, \mathbf{e}_f)\bigr)}
{\sum_{d \in \mathcal{D}_q^P} \exp\bigl(s_d(q, d, \mathbf{e}_f)\bigr)},
\label{eq:training_rank_loss}
\end{equation}
where $s_d(q, d, \mathbf{e}_f)$ is the FFN-predicted relevance score and $d^{*} \in \mathcal{D}_q^P$ is the target document. During training, the target document is guaranteed to appear in $\mathcal{D}_q^P$ via teacher forcing.

\textbf{Pairwise ranking loss.}
To additionally supervise relative ordering between different relevance levels---e.g., a strongly relevant document should outrank a weakly relevant one---we add a pairwise margin loss over all positive--negative pairs:
\begin{equation}
\mathcal{L}_{\mathrm{pair}}
=
\sum_{d^{+}\in\mathcal{P}_q}
\sum_{d^{-}\in\mathcal{N}_q}
\max\!\left(0,\;
\gamma - s_d(q,d^{+},\mathbf{e}_f) + s_d(q,d^{-},\mathbf{e}_f)
\right),
\label{eq:training_pairwise_loss}
\end{equation}
where $\gamma$ is the ranking margin.

\textbf{Joint objective.}
The complete training objective combines all three losses:
\begin{equation}
\mathcal{L}
=
\lambda_c \mathcal{L}_{\mathrm{cluster}}
+
\lambda_r \mathcal{L}_{\mathrm{rank}}
+
\lambda_p \mathcal{L}_{\mathrm{pair}},
\label{eq:training_total_loss}
\end{equation}
where $\lambda_c$, $\lambda_r$, and $\lambda_p$ balance the contributions of cluster-level coverage, document identification, and pairwise document ordering, respectively.

\subsection{Positive and Hard-Negative Mining}
\label{sec:sample_mining}

\subsubsection{Positive Sample Sources}

We construct positive samples from multiple supervision sources in order to capture different aspects of query--document relevance. The main sources include:

\begin{itemize}
    \item \textbf{Relevance supervision}: documents receiving high relevance grades from human annotators or teacher models, providing direct evidence of whether a document satisfies the information need;
    \item \textbf{Satisfaction supervision}: documents associated with high user satisfaction signals in actual search sessions;
    \item \textbf{Behavioral supervision}: documents associated with strong user consumption signals, such as long dwell time, effective clicks, or deep reading behavior.
\end{itemize}

Combining these sources reduces dependence on any single annotation or behavioral signal and improves coverage across different query types. For a query $q$, the resulting positive set is
\begin{equation}
\mathcal{P}_q
=
\mathcal{P}^{\mathrm{rel}}_q
\cup
\mathcal{P}^{\mathrm{sat}}_q
\cup
\mathcal{P}^{\mathrm{beh}}_q.
\label{eq:positive_union}
\end{equation}
When the same document appears in multiple sources, it is deduplicated and its supervision signals are combined according to the corresponding labeling policy.

\subsubsection{Negative Sample Sources}

We construct negative samples primarily from documents associated with the same query but having lower relevance labels or weaker user feedback. These samples are more informative than randomly selected documents because
they are often topically related to the query while failing to satisfy its specific information need.

The main negative sources include:

\begin{itemize}
    \item \textbf{Cross-grade relevance samples}: documents associated with the same query but assigned lower relevance grades than the positive documents, providing explicit supervision about relative document quality;
    \item \textbf{Click-based samples}: documents exposed for the query but receiving weak or no effective click signals, particularly when other documents for the same query receive stronger positive feedback.
    Click signals are interpreted together with exposure, position, and other contextual information, rather than treating every non-clicked document as definitively irrelevant;
    \item \textbf{Retrieval hard negatives}: documents retrieved with high scores by a retrieval system but judged by human annotators or teacher models to be irrelevant or insufficiently relevant.
\end{itemize}

The negative set is written as
\begin{equation}
\mathcal{N}_q
=
\mathcal{N}^{\mathrm{rel}}_q
\cup
\mathcal{N}^{\mathrm{click}}_q
\cup
\mathcal{N}^{\mathrm{retrieval}}_q,
\label{eq:negative_union}
\end{equation}
where $\mathcal{N}^{\mathrm{rel}}_q$ contains lower-relevance documents, $\mathcal{N}^{\mathrm{click}}_q$ contains documents with weak click-based evidence, and $\mathcal{N}^{\mathrm{retrieval}}_q$ contains retrieved but
insufficiently relevant documents.

These hard negatives are mixed into the candidate set $\mathcal{D}_q^P$ during training, encouraging the model to learn fine-grained distinctions between documents that share similar topics but differ in their ability to
satisfy the query intent.

\subsection{Rolling Training}
\label{sec:dynamic_training}

The query--document relevance distribution evolves continuously, driven by newly published documents, emerging topics, and shifting user behavior. A model trained on a fixed snapshot therefore degrades over time. We address
this with a rolling training strategy.

At each training round, newly collected queries, documents, relevance labels, and user-feedback signals are merged with a curated pool of historical high-quality samples, and used to continue training from the
previous checkpoint. The recent samples let the model adapt to new intents, new content, and new behavioral patterns, while the historical samples preserve long-tail relevance knowledge and mitigate catastrophic forgetting.
This design pairs explicit relevance supervision with implicit behavioral evidence in every round, keeping the model both accurate and timely.

\section{System Design}
\label{sec:system_design}

Deploying WeSearch as the primary retrieval path of a billion-scale production
search system requires more than a working model. In this section, we describe
five system components that together make the model usable in production:
(i)~an online retrieval architecture that places WeSearch alongside a single
lexical path (Sec.~\ref{sec:online_architecture}); (ii)~a rolling mounting
mechanism that makes freshly published documents searchable within minutes
(Sec.~\ref{sec:rolling_mounting}); (iii)~a periodic refresh procedure that
upgrades the model and the full index as a single atomic unit
(Sec.~\ref{sec:periodic_refresh}); (iv)~an ADC-based scoring kernel that
drives the document-level stage at low cost (Sec.~\ref{sec:wesearch_adc_system});
and (v)~system-level training and inference acceleration
(Sec.~\ref{sec:system_acceleration}).

\subsection{Online Retrieval Architecture}
\label{sec:online_architecture}

Figure~\ref{fig:wesearch_online_architecture} compares the existing retrieval
system with the proposed architecture. The existing system combines multiple
lexical and dense retrieval sources in parallel, whose candidates are merged
and passed to the downstream ranking module.

In the proposed architecture, WeSearch serves as the \emph{primary} semantic
retrieval path: given a query, it directly produces document candidates by
first selecting relevant clusters and then ranking atomic documents within
the resulting candidate space, consolidating multiple parallel dense and
lexical retrieval paths into a single end-to-end model. We retain one
syntax tree as a complementary lexical path for exact matching that
semantic retrieval cannot cover. The two paths run in parallel, and their
candidates are merged and deduplicated before being passed to the
downstream ranker that produces the final results.

\subsection{Rolling Mounting of New Documents}
\label{sec:rolling_mounting}

\begin{figure}[t]
    \centering
    \includegraphics[width=0.9\columnwidth]{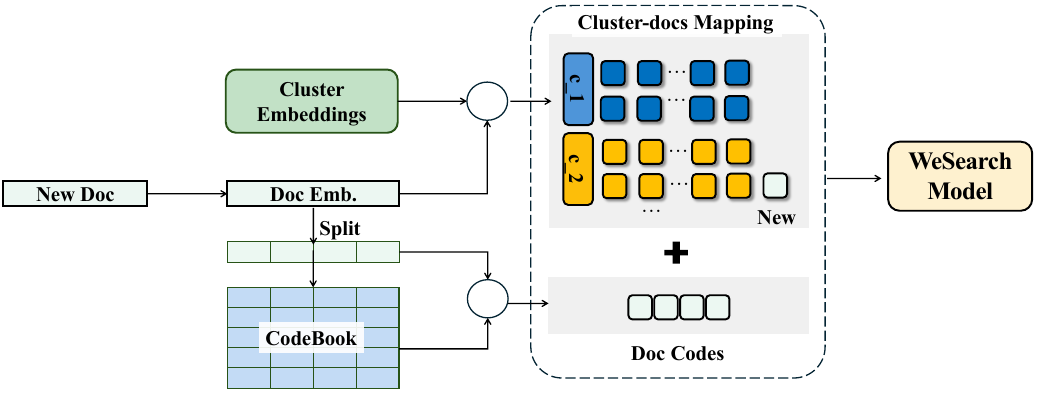}
    \caption{Rolling mounting of fresh documents. New documents are assigned to the existing cluster space, quantized with fixed codebooks, and made visible through an atomic index update.}
    \label{fig:rolling_mounting}
\end{figure}

Content-oriented search systems continuously receive newly published documents. Rebuilding the entire index for every new document is computationally expensive and introduces unnecessary delays before fresh content becomes searchable.

WeSearch supports incremental ingestion because clusters serve as candidate-space constraints rather than complete hierarchical document identifiers. A newly arriving document can therefore be assigned to the existing cluster space without modifying any model parameters. As shown in Figure~\ref{fig:rolling_mounting}, the rolling-mounting operation involves three steps: (1) computing the document representation and assigning it to the nearest cluster, (2) quantizing the representation with the fixed OPQ codebooks, and (3) appending the document and
its code to the cluster-to-document mapping.

Formally, for a new document $d_{\mathrm{new}}$, its encoder representation $\mathbf{h}_{\mathrm{new}} = f_{\theta}(d_{\mathrm{new}})$ is computed and assigned to the nearest cluster:
\begin{equation}
c_{\mathrm{new}}
=
\arg\min_{c\in\mathcal{C}}
\left\|
\mathbf{h}_{\mathrm{new}} - \mathbf{e}_c
\right\|_2.
\label{eq:new_document_cluster}
\end{equation}
The document is then quantized using the existing codebooks to obtain its compact code $\mathbf{z}_{\mathrm{new}} = Q(\mathbf{R}\,\mathbf{h}_{\mathrm{new}})$, and the cluster-to-document mapping is updated as $\mathcal{D}_{c_{\mathrm{new}}} \leftarrow \mathcal{D}_{c_{\mathrm{new}}} \cup \{d_{\mathrm{new}}\}$.

Because the cluster centroids and codebooks remain fixed during mounting, the operation is lightweight and can be performed continuously in the background, making fresh documents searchable with minimal latency.

\subsection{Periodic Model and Index Refresh}
\label{sec:periodic_refresh}
We do not rebuild the cluster–document mapping after every training update. Instead, model training and index refresh are managed as separate processes. When a new retrieval-model version is selected for deployment, we refresh the associated index artifacts as needed: document-to-cluster assignments are recomputed using the document representations and cluster embeddings of that version, and document codes are generated using its matching OPQ parameters. The model and corresponding cluster embeddings, OPQ parameters, document codes, and cluster–document mapping are validated and released as a consistent version. Between such refreshes, newly arriving documents are mounted into the active index using the current version’s fixed cluster space and OPQ parameters.

\subsection{ADC Optimization for Online Document Scoring}
\label{sec:wesearch_adc_system}

During online serving, the candidate set $\mathcal{D}_q$ can be very large. Explicitly reconstructing a full document vector for every candidate and computing full-dimensional distances would incur substantial computation and memory-access cost.

We instead use Asymmetric Distance Computation (ADC)~\cite{10.1109/TPAMI.2010.57} to score compressed document
representations directly. Before scoring any candidate, the rotated query $\widetilde{\mathbf{h}}_q = \mathbf{R}\mathbf{h}_q$ is partitioned into $m$ sub-vectors $\{\widetilde{\mathbf{h}}^{(j)}_q\}_{j=1}^{m}$, and a
query-dependent lookup table is constructed for each sub-space:
\begin{equation}
\operatorname{LUT}_j[k]
=
\left\|\widetilde{\mathbf{h}}^{(j)}_q - \mathbf{b}_{j,k}\right\|_2^2,
\qquad k=1,\ldots,K.
\label{eq:system_lookup_table}
\end{equation}
The approximate squared L2 distance between the query and a document with code $\mathbf{z}_d = [z^{(1)}_d,\ldots,z^{(m)}_d]$ is then obtained by table lookups and additions:
\begin{equation}
\widehat{\delta}(q,d)
=
\sum_{j=1}^{m}\operatorname{LUT}_j\!\left[z^{(j)}_d\right].
\label{eq:system_adc_distance}
\end{equation}
Ranking by $-\widehat{\delta}(q,d)$ is order-equivalent to ranking by $s_{\text{coarse}}(q,d)$ in Sec.~\ref{sec:wesearch_document_decoding}, so ADC provides a drop-in acceleration of the coarse-ranking stage. Each lookup table is built once per query and reused across all candidates and across distributed partitions; document-side computation reduces to lookups and additions rather than full-dimensional multiplications, substantially cutting compute and memory traffic while retaining an accurate approximation of the semantic distance.
\subsection{System-Level Acceleration}
\label{sec:system_acceleration}

We introduce several system-level optimizations to improve the efficiency of WeSearch training and online inference, targeting distributed document-code access, tensor-parallel execution~\cite{shoeybi2020megatronlmtrainingmultibillionparameter}, communication overhead, and kernel performance.

\subsubsection{Training Acceleration}
\label{sec:training_acceleration}

Document codes are partitioned across devices by cluster membership in a load-balanced fashion, keeping each partition local to its assigned device. This enables tensor-parallel computation at the document level without hosting the full document-code vocabulary on a single device.
To reduce cross-device communication, only query representations and query-associated states are exchanged between devices; document codes remain resident locally. Each device independently accesses its local partition while
required query embeddings are broadcast among tensor-parallel workers, greatly reducing communication volume during document-space construction and score computation.

Tensor-parallel communication is overlapped with local computation: while transmitting query representations or intermediate states, each device processes document codes already available locally, lowering synchronization cost and improving hardware utilization. At the kernel level, we optimize distributed cross-entropy computation to reduce the cost of global normalization and logit synchronization across devices.

\subsubsection{Inference Acceleration}
\label{sec:inference_acceleration}

For online serving, WeSearch is integrated with the vLLM inference engine~\cite{kwon2023efficient} to accelerate query encoding, delivering optimized scheduling, batching, and execution under live traffic.

We further reduce serialization and inter-process communication overhead within the serving pipeline. Intermediate results are transferred using compact representations, and unnecessary serialization, deserialization, and memory-copy operations are eliminated. Candidate identifiers, scores, quantized codes, and document-side features are transmitted only when required by downstream stages, reducing communication latency and improving throughput for candidate merging, deduplication, and final result generation.

\section{Experiments}
\label{sec:experiments}

We evaluate WeSearch in three stages. First, we conduct controlled experiments on the Base corpus to examine fundamental design choices, including cluster granularity, hierarchical Semantic-ID-based retrieval, and document
representation. Second, we evaluate the effect of structured document features and training-data mining on the Expanded corpus. Third, we assess the complete retrieval architecture through online testing, including deployment with fresh-document ingestion.

The experiments are designed to answer the following questions:
\begin{enumerate}
    \item \textbf{Retrieval design}: How do cluster granularity, hierarchical Semantic IDs, and document representation affect retrieval quality?
    \item \textbf{Training signals}: Do structured document features and multi-source training-data mining improve retrieval quality across different query subsets?
    \item \textbf{Production utility}: Does the complete WeSearch architecture, combined with a complementary lexical syntax tree, improve result quality and online search performance?
\end{enumerate}

\subsection{Experimental Setup}
\label{sec:experimental_setup}

\subsubsection{Search Corpus}
\label{sec:search_corpus}

We evaluate WeSearch in the WeChat Search scenario. The corpus contains multimodal image-text documents---including articles and graphic-text posts---together with document metadata and user interaction signals. We consider three corpus configurations:
\begin{itemize}
    \item \textbf{Base corpus}: hundreds of millions of documents, used for controlled feasibility studies and ablation experiments;
    \item \textbf{Expanded corpus}: billions of documents, obtained by extending the Base corpus with additional relatively stable document collections;
    \item \textbf{Expanded corpus with Fresh corpus}: the Expanded corpus augmented with a newly arriving corpus of tens of millions of documents.
\end{itemize}

\subsubsection{Evaluation Data}
\label{sec:evaluation_data}

We use two types of offline evaluation data: (1) randomly sampled queries with human relevance annotations, and (2) queries grouped by user-consumption signals, including long dwell time and effective clicks. For teacher-based evaluation, we retrieve the top-$100$ documents per query and assign quality labels using a satisfaction teacher and a relevance teacher. A document is considered high-quality if it receives a high-quality grade from the corresponding teacher.

\subsubsection{Metrics}
\label{sec:metrics}

For a query $q$, let $\mathcal{G}_q$ be the set of ground-truth documents and $\mathcal{R}^{@k}_q$ the top-$k$ retrieved documents. Recall at $k$ is defined as
\begin{equation}
\operatorname{Recall}@k
=
\frac{
\left|\mathcal{R}^{@k}_q \cap \mathcal{G}_q\right|
}{
\left|\mathcal{G}_q\right|
}.
\label{eq:experiment_recall}
\end{equation}

For teacher-based evaluation, let $\mathbb{I}_{\mathrm{hq}}(q,d)$ indicate whether document $d$ receives a high-quality label. The high-quality result density is defined as
\begin{equation}
\operatorname{Density}_{\mathrm{hq}}
=
\frac{1}{|\mathcal{Q}|}
\sum_{q\in\mathcal{Q}}
\frac{
\sum_{d\in\mathcal{R}^{@K}_q}
\mathbb{I}_{\mathrm{hq}}(q,d)
}{K}.
\label{eq:good_result_density}
\end{equation}
We report satisfaction density and relevance density according to the corresponding teacher models.

\subsection{Compared Methods}
\label{sec:compared_methods}

We compare the following retrieval methods:
\begin{itemize}
    \item \textbf{Online baseline}: the previous production configuration, combining inverted-index retrieval with dense vector retrieval. 
    \item \textbf{Semantic-ID generation}: a hierarchical generative retriever that represents each document with a fixed Semantic ID and decodes it
    level by level;
    \item \textbf{WeSearch}: the proposed two-stage model-centric retriever, which
    first selects clusters and then ranks atomic documents within the
    query-specific candidate space;
    \item \textbf{WeSearch + one syntax tree}: the complete retrieval architecture,
    combining WeSearch with one lexical syntax tree.
\end{itemize}

Semantic-ID-based retrieval is used only in the controlled Base-corpus experiments. The online baseline, WeSearch, and WeSearch with one syntax tree are evaluated in the large-scale and online experiments. 

\subsection{Base-Corpus Feasibility Study}
\label{sec:base_corpus_study}

\begin{figure}[t]
    \centering
    \includegraphics[width=0.95\columnwidth]{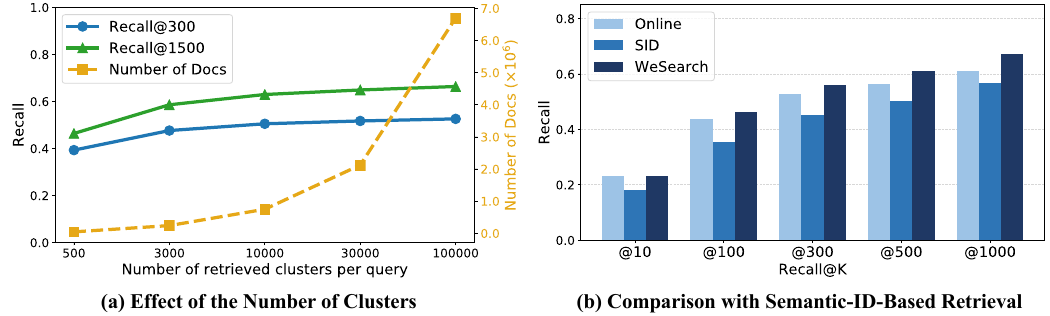}
    \caption{Experimental results on the Base corpus.}
    \label{fig:exp_fig1}
\end{figure}

We conduct controlled experiments on the Base corpus using an in-distribution test set. The study examines three design questions: the number of clusters retrieved per query, the suitability of hierarchical Semantic-ID-based retrieval for graphic-text documents, and the impact of document representation and quantization configuration.

\subsubsection{Effect of the Number of Retrieved Clusters Per Query}
\label{sec:cluster_number_study}

We vary the number of clusters retrieved per query, denoted by $N$. As shown in Figure~\ref{fig:exp_fig1}(a), increasing $N$ initially improves candidate coverage, but gains become marginal once $N$ reaches approximately 10,000, while inference cost continues to increase. We therefore set $N=10{,}000$ in subsequent experiments.

\subsubsection{Comparison with Semantic-ID-Based Retrieval}
\label{sec:base_sid_study}

Figure~\ref{fig:exp_fig1}(b) compares WeSearch with the hierarchical Semantic-ID-based generative retriever and the online baseline. The Semantic-ID-based method underperforms both the online baseline and WeSearch at
all recall cutoffs, while WeSearch consistently outperforms the online baseline.
These results indicate that Semantic-ID-based retrieval is not well suited to content-oriented search over graphic-text documents, where cluster-based candidate reduction followed by atomic document ranking is more effective.

\subsubsection{Document Representation and Quantization}
\label{sec:base_quantization_study}

\begin{table}[htbp]
\centering
\resizebox{0.45\linewidth}{!}{
  \begin{tabular}{ccc}
    \hline
    Clustering & Doc Memory & Recall@300 \\
    \hline
    $2k\times1k$   & OPQ$(32, 256)$ & 46.66\% \\
    $2k\times1k$   & OPQ$(32, 512)$ & 47.10\% \\
    $2k\times1k$   & OPQ$(64, 256)$ & 51.54\% \\
    $10k\times200$ & OPQ$(64, 256)$ & 51.34\% \\
    $2k\times1k$   & Dense($32$)    & 33.07\% \\
    \hline
  \end{tabular}
  }
  \caption{Ablation on clustering structure and document memory. $a\times b$ denotes two-level RQ-KMeans with $a$ coarse clusters, each split into $b$ sub-clusters. $\text{OPQ}(m, K)$ denotes $m$ sub-spaces
  with codebook size $K$ (Sec.~\ref{sec:wesearch_compression}); Dense($32$) stores a $32$-dim embedding without quantization. Cluster Recall@$10k$
  is stable across all rows and is therefore omitted.}
  \label{tab:ablation}
\end{table}

Table~\ref{tab:ablation} reports the effect of document representation and quantization under a fixed cluster setting. Increasing the codebook size from $256$ to $512$ improves document-level recall, and increasing the number of OPQ subvectors from $32$ to $64$ provides a further gain, indicating that both codebook capacity and representation partitioning affect document discrimination.
Varying the cluster-count split between the two RQ-KMeans layers has only a limited effect, suggesting that retrieval quality is less sensitive to this particular clustering configuration than to the capacity of the document representation. Finally, OPQ$(64,256)$ substantially outperforms the $32$-dimensional dense representation, confirming that directly reducing the embedding to a low-dimensional vector discards discriminative information, whereas OPQ preserves it by partitioning the representation into multiple
subspaces.

\subsection{Large-Scale Validation}
\label{sec:large_scale_validation}

We evaluate WeSearch on the Expanded corpus, focusing on the contribution of structured document features and training-data mining.

\subsubsection{Document Features and Training-Data Mining}
\label{sec:feature_and_mining_validation}

\begin{table}
    \centering
    \resizebox{0.8\columnwidth}{!}{
    \begin{tabular}{lcccc}
        \toprule
        Method
        & Relevance--1
        & Relevance--2
        & Long Dwell
        & Effective Click \\
        \midrule
        Online
        & 0.3621 & 0.3087 & 0.4407 & 0.4604 \\
        \midrule
        WeSearch w/o features \& mining
        & 0.4072 & 0.3277 & 0.4323 & 0.4379 \\
        WeSearch w/o mining
        & 0.4271 & \textbf{0.3865} & 0.4961 & 0.4973 \\
        WeSearch (full)
        & \textbf{0.4344}
        & 0.3794
        & \textbf{0.5505}
        & \textbf{0.5481} \\
        \bottomrule
    \end{tabular}}
    \caption{Ablation of structured document features and training-data mining in WeSearch, measured by Recall@100. ``w/o features \& mining''
    uses only semantic embeddings and the basic relevance supervision;
    ``w/o mining'' adds structured document features; the full model
    further incorporates multi-source positive/negative mining.}
\label{tab:feature_mining_ablation}
\end{table}

We ablate the two training-signal interventions in WeSearch by progressively removing multi-source training-data mining and structured document features from the full model, and compare each variant with the online baseline on
four query subsets: two relevance-oriented sets (Relevance--1, Relevance--2) and two behavior-oriented sets (Long Dwell, Effective Click). Recall@100 is reported in Table~\ref{tab:feature_mining_ablation}.

Even without features and mining, WeSearch already surpasses the online baseline on both relevance-oriented subsets, confirming that the two-stage cluster--then-document design itself is a strong semantic retriever.
Adding structured document features yields consistent gains across all four subsets, including a substantial lift on the two behavior-oriented subsets, confirming that document-side quality, user-behavior, and temporal
attributes provide signals complementary to pure content embeddings.
Further adding multi-source positive/negative mining achieves the best overall results on Relevance--1, Long Dwell, and Effective Click, with the largest gains concentrated on the behavior-oriented subsets---matching the
design intent of combining relevance, satisfaction, and behavioral supervision to discriminate documents that are topically relevant but differ in their ability to satisfy user intent.

\subsubsection{Teacher-Assessed Result Quality}
\label{sec:teacher_quality}

\begin{table}[t]
    \centering
    \begin{tabular}{lcc}
        \toprule
        Method
        & Satisfaction density
        & Relevance density \\
        \midrule
        Online
        & 0.1732 & 0.4575 \\
        WeSearch
        & 0.1871 & 0.4662 \\
        WeSearch + one syntax tree
        & \textbf{0.2088} & \textbf{0.4854} \\
        \bottomrule
    \end{tabular}
    \caption{Teacher-assessed high-quality result density.}
    \label{tab:teacher_density}
\end{table}

Table~\ref{tab:teacher_density} reports satisfaction density and relevance density for the production baseline, WeSearch, and WeSearch with one syntax tree. WeSearch improves both metrics over the production baseline. Adding one syntax tree provides further gains, indicating that the syntax tree contributes lexical coverage complementary to WeSearch's semantic retrieval. These results support the design in which WeSearch serves as the primary semantic retrieval path while the syntax tree is retained as a complementary lexical source.

\subsection{Online Evaluation}
\label{sec:online_evaluation}

We deploy WeSearch as the primary retrieval path in the WeChat Search pipeline, retaining a single syntax-tree branch as a complementary lexical source. Candidates from both branches are merged and deduplicated before
entering the downstream ranking module. The deployed system is assessed along four complementary dimensions: a one-week online A/B test, a one-month online A/B test, pairwise human evaluation, and measurements of serving
cost and latency.

\subsubsection{One-Week Online A/B Test}
\label{sec:one_week_online_results}

We first conduct a one-week online A/B test against the corresponding production baseline under each of the three corpus configurations.
Table~\ref{tab:one_week_online_results} reports the relative changes in CTR, together with the results on long-query traffic.

\begin{table}[t]
    \centering
    \resizebox{0.5\columnwidth}{!}{
    \begin{tabular}{lc}
        \toprule
        Evaluation group & CTR \\
        \midrule
        Base corpus & \textbf{+0.27\%} \\
        Expanded corpus & \textbf{+0.61\%} \\
        Expanded corpus with Fresh corpus & \textbf{+0.53\%} \\
        \midrule
        Long-query traffic$^{\dagger}$ & \textbf{+1.10\%} \\
        \bottomrule
    \end{tabular}
    }
    \caption{Relative changes in CTR in the one-week online A/B test, compared with the corresponding production baselines. The first three rows report full-traffic results under each corpus configuration.
    $^{\dagger}$Long-query traffic is evaluated on the Expanded corpus with Fresh corpus configuration. Bold values indicate statistically
    significant gains ($p<0.05$).}
    \label{tab:one_week_online_results}
\end{table}

WeSearch yields consistent CTR improvements across all three corpus configurations, with gains ranging from $+0.27\%$ to $+0.61\%$. The improvement extends to the long-query segment, where CTR increases by
$+1.10\%$, suggesting that WeSearch benefits not only aggregate traffic but also the more challenging long-tail queries where semantic understanding is critical.

\subsubsection{One-Month Online A/B Test}
\label{sec:one_month_online_results}

To verify that these gains persist over a longer horizon, we conduct a one-month full-traffic A/B test on the Expanded corpus with Fresh corpus, comparing WeSearch with the corresponding production baseline.
Table~\ref{tab:one_month_online_results} reports the relative changes in search metrics as well as per-user article-engagement metrics. All reported improvements are statistically significant ($p<0.05$).

\begin{table}[t]
    \centering
    \begin{tabular}{lc}
        \toprule
        Metric & Relative change \\
        \midrule
        CTR & \textbf{+0.68\%} \\
        Dwell time$^{\dagger}$ & \textbf{+0.71\%} \\
        Article clicks$^{\ddagger}$ & \textbf{+2.27\%} \\
        \bottomrule
    \end{tabular}
    \caption{Relative changes in the one-month online A/B test on the Expanded corpus with Fresh corpus under full traffic, compared with the corresponding production baseline.
    $^{\dagger}$Average per-user dwell time on the search results page.
    $^{\ddagger}$Per-user clicks on WeChat Official Account articles from
    search results. Bold values indicate statistically significant gains
    ($p<0.05$).}
    \label{tab:one_month_online_results}
\end{table}

Over the one-month evaluation, CTR improves by $+0.68\%$ and the average dwell time on the search results page increases by $+0.71\%$, confirming that the one-week gains are stable rather than transient and that users
spend more time engaging with the returned results. More notably, per-user article clicks increase by $+2.27\%$, substantially exceeding the aggregate CTR gain. This indicates that WeSearch not only improves per-impression click quality but also drives considerably higher user-level engagement with WeChat Official Account articles, which is a key business objective of the search platform.

\subsubsection{Pairwise Human Evaluation}
\label{sec:gsb_evaluation}

To complement the behavioral metrics, we conduct blind pairwise human evaluations under all three corpus configurations, in which annotators compare the result lists of WeSearch and the previous production system
side by side. WeSearch achieves positive Better rates of $2\%$, $3\%$, and $3\%$ on the Base corpus, Expanded corpus, and Expanded corpus with Fresh corpus, respectively. The uniformly positive Better rates provide
direct human-perceived evidence of result-quality improvement that is consistent with the gains observed in the online A/B tests.

\subsubsection{Latency and Resource Cost}
\label{sec:latency_resource_cost}

Finally, we measure the system-level serving cost and latency against the previous production configuration. The proposed architecture reduces overall retrieval resource cost by $41\%$ and lowers graphic-text
retrieval latency by $19\,\mathrm{ms}$. Together with the quality gains reported above, these results demonstrate that WeSearch delivers improvements in online search metrics while simultaneously reducing retrieval-stage resource consumption and latency, which is critical for large-scale production deployment.

\section{Conclusion}
\label{sec:conclusion}

We presented \textbf{WeSearch}, a model-centric retrieval framework for large-scale, continuously evolving content-oriented search. WeSearch uses clusters to reduce the query-specific candidate space and performs
fine-grained matching and ranking over atomic documents, without assigning documents fixed hierarchical Semantic IDs. Its production architecture combines OPQ-based document compression, ADC-accelerated candidate scoring,
and rolling mounting for fresh-document ingestion. In WeChat Search, WeSearch serves as the primary semantic retrieval path, complemented by a lexical syntax tree for exact term and phrase matching.

We evaluated WeSearch across corpus configurations ranging from hundreds of millions to billions of documents. One-week online A/B tests show consistent CTR improvements across all three configurations, with gains
further confirmed by a one-month full-traffic A/B test and blind pairwise human evaluation. On the largest configuration, WeSearch improves CTR by $0.68\%$, per-user article clicks by $2.27\%$, and average dwell time on
the search result page by $0.71\%$, while reducing overall retrieval resource cost by $41\%$ and graphic-text retrieval latency by $19$\,ms. WeSearch has been fully deployed in WeChat Search, serving $100\%$ of
production traffic, demonstrating the effectiveness and production scalability of cluster-constrained retrieval for large-scale content search.

Several directions remain for future work. First, incorporating user profiles, session context, and real-time behavioral signals could enable personalized cluster selection and document ranking while controlling
serving cost. Second, tighter integration of candidate retrieval and downstream reranking, for example through shared representations and coordinated training objectives, may improve consistency between candidate generation and final ranking.

\bibliography{main}
\bibliographystyle{plainnat}

\end{document}